\documentclass[a4paper,11pt]{article}
\pdfoutput=1 % if your are submitting a pdflatex (i.e. if you have
\usepackage{jcappub} % for details on the use of the package, please
\usepackage{graphicx}
\usepackage{amsmath}
\usepackage{amssymb}
\usepackage[all]{hypcap}
\usepackage{mathptmx}
\usepackage{txfonts}
\usepackage{ae,aecompl}
\usepackage[normalem]{ulem}
\usepackage{longtable}
\usepackage{multirow}
\usepackage{multicol}
\usepackage{tablefootnote}
\usepackage{appendix}
\usepackage{xcolor}
\usepackage{pagecolor}
\usepackage{hyperref}
\usepackage{chngcntr}
\usepackage{orcidlink}

\definecolor{hcolo}{RGB}{0,0,190}
\hypersetup{colorlinks=true,linkcolor=hcolo,citecolor=hcolo,filecolor=hcolo,urlcolor=hcolo}
\def\mpc{\,{\rm Mpc}}

\def\kpsm{\,{\rm km\, s^{-1} \, Mpc^{-1}}}

\title{Unveiling galaxy pair alignment in cosmic filaments: A 3D exploration using EAGLE simulation}

\author[a]{Suman Sarkar\,\orcidlink{0000-0002-5465-3467}}
\author[b]{, Biswajit Pandey\,\orcidlink{0000-0001-7876-595X}}

\affiliation[a]{{\small Department of Physics, Indian Institute of Technology Kharagpur, West Bengal - 721302, India.}}
\affiliation[b]{{\small Department of Physics, Visva-Bharati University, Santiniketan, West Bengal - 731235, India.}}

\emailAdd{suman2reach@gmail.com}
\emailAdd{biswap@visva-bharati.ac.in}

\abstract{We investigate how galaxy pairs are oriented in three
  dimensions within cosmic filaments using data from the EAGLE
  simulation.  We identify filament spines using DisPerSE and isolate
  galaxies residing in filamentary environments. Employing a FoF
  algorithm, we delineate individual filaments and determine their
  axes by diagonalizing the moment of inertia tensor. The orientations
  of galaxy pairs relative to the axis of their host filament are
  analyzed. Our study covers diverse subsets of filaments identified
  through varying linking lengths, examining how galaxy pairs align
  with the filament axis across different spatial parameters such as
  pair separation and distance from the filament
  spine. \textcolor{black}{We observe a nearly uniform probability
    distribution for the cosine of the orientation angle, which is
    nearly identical in each case.  We also investigate the effects of
    redshift space distortions and confirm that the probability
    distributions remain uniform in both real space and redshift
    space. To validate our approach, we conduct Monte Carlo
    simulations using various theoretical probability
    distributions. Our analysis does not reveal any evidence of
    preferential alignment of galaxy pairs within cosmic filaments in
    hydrodynamical simulations.}}


\begin{document}
\maketitle
\flushbottom
\section{Introduction}\label{sec:intro}

Galaxies, the building blocks of the universe, are known to be
distributed along a complex, interconnected network of clusters,
filaments and sheets surrounded by vast empty regions. This network of
galaxies is often referred to as the ``cosmic web'' \cite{bond96}.
The galaxies and their host dark matter halos residing in different
cosmic web environments, experience the tidal force from the
surrounding large-scale structures (LSS). Various observations suggest
that the tidal force from the LSS can influence the orientation of the
galaxies \cite{lambas88, lee02, donoso06, lee07, hirv17}. It has been
reported that the major axes of the galaxies in filaments and sheets
align preferentially along the filament axes and the plane of the
sheet respectively \cite{jones10, zhang13, tempel13a, tempel13b,
  zhang15, chen19, krolewski19}. The dark matter halos in cosmological
simulations are also known to have specific alignment with respect to
their host cosmic web environment \cite{aragon07, zhang09,
  libeskind12, ganeshaiah18}.

Besides the alignment of the galaxies and their host dark matter
halos, the alignment of the galaxy pairs with their host filaments has
been also reported in the literature \cite{tempel15a, mesa18}. The
filaments are thread-like structures that connect the galaxy clusters
\cite{einasto80, pimbblet05}. They represent the denser parts of the
cosmic web and host $\sim 40\%-45\%$ of the total mass in the universe
\cite{aragon10, haider16}. A large number of close galaxy pairs are
located inside the filaments. The galaxies in pairs are interacting
gravitationally with their companions as well as with their embedding
large-scale environments. The preferential alignment of the galaxy
pairs with the filament axes may arise due to a complex interplay of
the gravitational forces operating on small and large-scales. Studying
such alignment can provide important insights into the dynamics of the
cosmic web. Further, both galaxy interactions \cite{larson78,
  barton00, lambas03, alonso06, ellison08a} and the large-scale
environments \cite{pandey06, pandey07, scudder12, luparello15,
  pandey17, lee18, maret18, pandey20, sarkar20, sarkar21} are known to
influence the galaxy properties and their evolution. Understanding the
alignment may also provide crucial information on the formation and
evolution of galaxies.

The galaxy pairs in observations are generally identified by applying
cuts on the projected separation and the velocity difference
\cite{alonso06, ellison08b, das1}. It is difficult to ensure that the
two members in such pairs are close in three dimension. Some of the
pairs may purely arise due to chance superposition in the high-density
regions \cite{alonso04}. Further, the galaxies are mapped in the
redshift space where the peculiar velocities distort the clustering
pattern. The redshift space distortions (RSD) is a major obstacle in
measuring the orientation of galaxy pairs with respect to the
filaments. Previous works \cite{tempel15a, mesa18} measure the
projected angle between the orientations of the filament and the
galaxy pairs. The filaments and the galaxy pairs are projected on the
plane of the sky to minimise the affects of RSD. They find $\sim
15\%-25\%$ excess of aligned pairs in filaments compared to a random
distribution. The total number of pairs associated to filaments in
these works are $3012$ and $4614$ respectively. The hydrodynamical
simulations can provide larger dataset of paired galaxies that can be
used to study such alignment. One can also quantify the effects of RSD
by comparing the alignment signals in real and redshift space.

\textcolor{black}{Any potential alignment signal highlights the
  significant influence of tidal fields from large-scale coherent
  structures on the spatial distribution of galaxies within these
  environments. Such alignment suggests a correlation between galaxy
  properties and their large-scale environments. When galaxy pairs
  align with filaments, it may indicate that their formation and
  evolution are influenced by tidal fields and gas flows along these
  structures. For example, this alignment could affect the efficiency
  of gas accretion in these galaxies. Recent studies by \cite{das2}
  analyze galaxy pairs in filaments and sheets using SDSS data and
  find that star formation rates in filaments are significantly higher
  than in sheets. These findings are further supported by \cite{das3},
  which uses data from EAGLE simulations.  Further, galaxy pairs
  oriented along cosmic filaments can have important implications for
  weak lensing studies in cosmology. The preferred alignment may
  significantly impact measurements of cosmic shear and weak lensing,
  influencing shear alignment, introducing systematic effects,
  affecting mass reconstruction, and suggesting environmental
  dependencies in weak lensing signals. Accurate modeling and analysis
  are essential to disentangle these effects and properly interpret
  weak lensing observations in cosmology.}

We aim to analyze data from the EAGLE simulations to measure the 3D
orientation of galaxy pairs within filaments. We will use the DIScrete
PERsistent Structures Extractor (DisPerSE) \cite{sousbie11a,
  sousbie11b} to identify galaxies residing in filamentary
environments. Using a Friend of Friend (FoF) algorithm
\cite{huchra82, berlind06}, we will separate these galaxies into
different filaments, resulting in a catalog of individual
filaments. Axis of each filament will be determined by diagonalizing
its moment of inertia tensor, with the primary axis represented by the
eigenvector corresponding to the smallest eigenvalue. Galaxy pairs
within each filament will be identified based on their 3D
separation. Both filaments and galaxy pairs will be identified in
three dimensions. The orientation angle between galaxy pairs and their
host filament axes will be measured for all pairs within
filaments. Any preferential alignment of galaxy pairs with respect to
filament axes will be reflected in the probability distribution of the
cosine of these orientation angles. The alignment could be influenced
by both the separation between the two members of a pair and the
distance of their center of mass from the filament spine. We will
investigate how the alignment signal varies with both the separation
between the two galaxies in each paired system and their distance from
the filament spine. Additionally, we will map simulated galaxies in
redshift space using their peculiar velocities and repeat the entire
analysis to assess the effects of RSD on these measurements.

The structure of the paper is outlined as follows: Section 2 describes
the data, Section 3 explains our methods, Section 4 discusses the
results of our analysis, and Section 5 presents our conclusions.

\section{Data}\label{sec:data}

\noindent The EAGLE project \cite{schaye15,crain15} consists of
large-scale hydrodynamical simulations of the Lambda-Cold Dark Matter
universe. These simulations have been crucial in gaining insights into
the behavior of dark matter and galactic evolution. The project adopts
the flat $\Lambda$CDM cosmology with $\Omega_\Lambda=0.693$,
$\Omega_m=0.307$, $\Omega_b=0.04825$, and $H_0=67.77\,\, \kpsm$
\cite{planck2014}. The simulations track the evolution of dark matter
and baryons in comoving cubes of sizes $25$, $50$, and $100$ Mpc using
an advanced version of GADGET2 \cite{springel05}, starting from
redshift 127 down to redshift 0. For this study, we focus on the final
snapshot ($Snapnum=28$) of the simulation at $z=0$, covering a cubic
volume of $100$ comoving Mpc per side. The Cartesian coordinates of
subhalos, represented by their center-of-potential, are extracted from
the $Ref-L0100N1504\_Subhalo$ table \cite{mcalpine16}. Galaxies linked
to these subhalos are identified using the $Ref-L0100N1504\_Aperture$
table, employing a spherical aperture of $30$ kpc and requiring that
their stellar mass is non-zero. To ensure that all unusual objects are
discarded, only galaxies with $Spurious=0$ are considered, resulting
in a total of $325358$ galaxies with unique galaxy IDs. We consider
only the galaxies with $\log(M_{\text{tot}} / M_{\odot}) \geq 8.5$,
giving us a sample consisting of $310759$ galaxies.
\textcolor{black}{For our analysis, we use a single data cube from the
  EAGLE simulation. We generate $100$ realizations from the original
  data cube by randomly selecting $80 \%$ of the entire population
  each time without repetition. The DisPerSE and FoF algorithms
  identify slightly different sets of filaments in each realization,
  which enables us to estimate the uncertainties in our
  measurements. Reducing the sampling rate to lower levels lead to
  larger differences in the filament identification and greater
  uncertainties in the measurement of their orientations.}  Finally,
we have $100$ data cubes, each containing $248607$ galaxies confined
within a $100 \mpc$ cube.

\section{Methods of analysis}\label{sec:methods}
\subsection{Finding galaxies in filaments using DisPerSE} \label{subsec:disperse}
\noindent To identify filament spines and nodes in the simulated
galaxy distribution from EAGLE, we utilize the DIScrete PERsistent
Structures Extractor (DisPerSE) \cite{sousbie11a,sousbie11b}. DisPerSE
is a scale-independent topological algorithm that delineates the
geometric features of the cosmic web components. It employs the
discrete Morse-Smale complex, formed by intersecting ascending and
descending manifolds of discrete Morse-Smale functions. The Morse
function is constructed using the Delaunay Tessellation Field
Extractor (DTFE) \cite{schaap00, weygaert09}, which transforms the
discrete galaxy distribution into a smooth density field. The
intersection of these manifolds reveals critical points corresponding
to maxima, minima, and saddle points in the density gradient. These
critical points are categorized by their indices 0, 1, 2, and 3,
representing voids, walls, filaments, and nodes, respectively. The
Morse field lines connect maxima and saddle points, with the points
along these lines indexed as -1, indicative of filament
spines. Structures exceeding a $>5 \sigma$ persistence ratio threshold
are retained to avoid identifying spurious patterns. Additionally,
galaxies within $d_s \leq 2 \mpc$ from the filament spine are
considered, where $d_s$ denotes the distance from the filament
spine. To exclude galaxies near nodes, a threshold of $d_n > 3 \mpc$
is applied, where $d_n$ is the distance from the nearest node with a
critical index of 3. \textcolor{black}{The choices for $d_n$ and $d_s$
  are somewhat subjective in our analysis. We primarily choose these
  values based on the typical radius of galaxy clusters and cosmic
  filaments. Our choices are somewhat larger than the typical radius
  of these structures to ensure that we exclude all galaxies from
  nodes and include all possible galaxy pairs within filaments.}

\begin{figure*}[ht!]%
\centering
\includegraphics[width=0.45\textwidth, trim=0.0cm 0.0cm 0.0cm 0.0cm, clip=true]{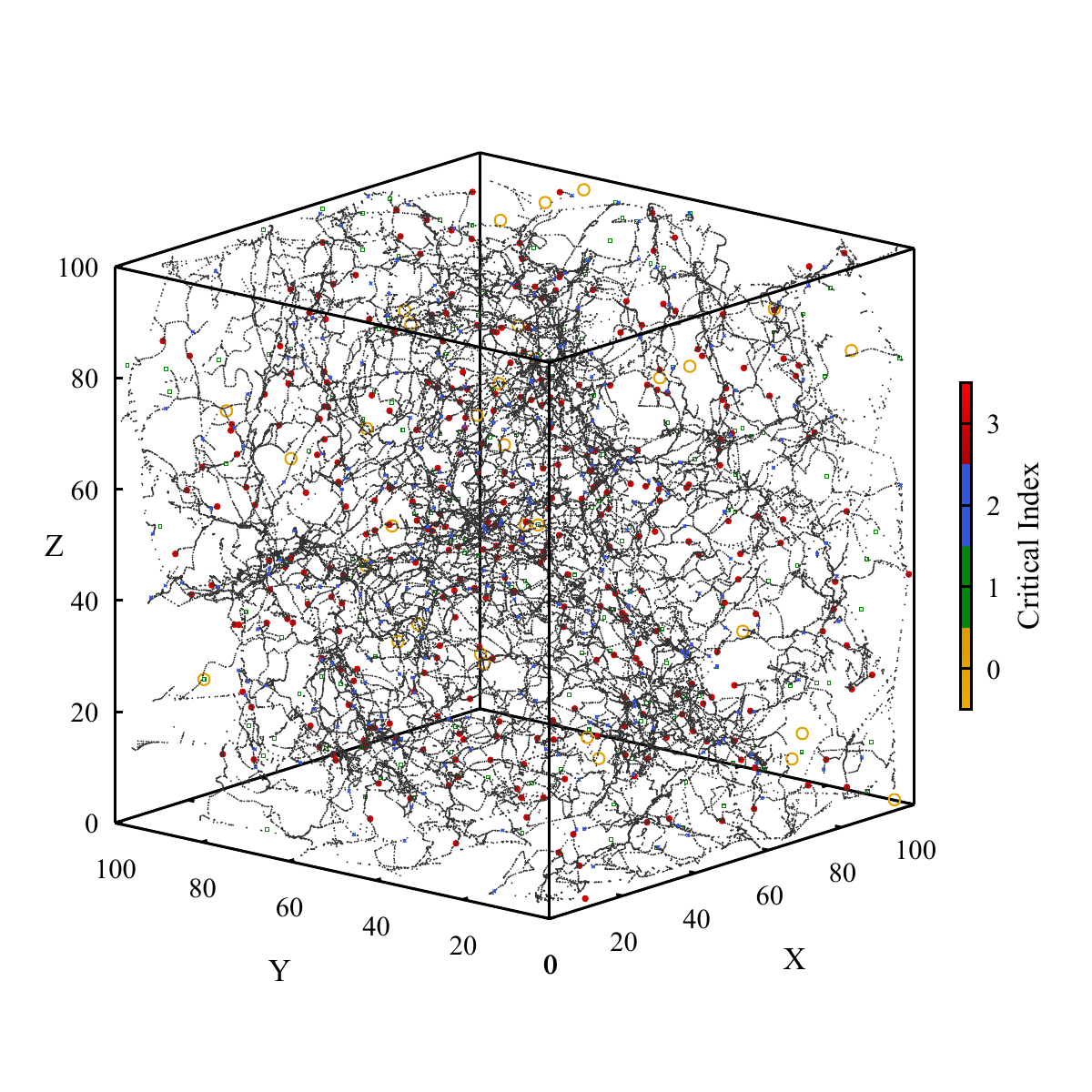} \hspace*{5px}
\includegraphics[width=0.45\textwidth, trim=0.0cm 0.0cm 0.0cm 0.0cm, clip=true]{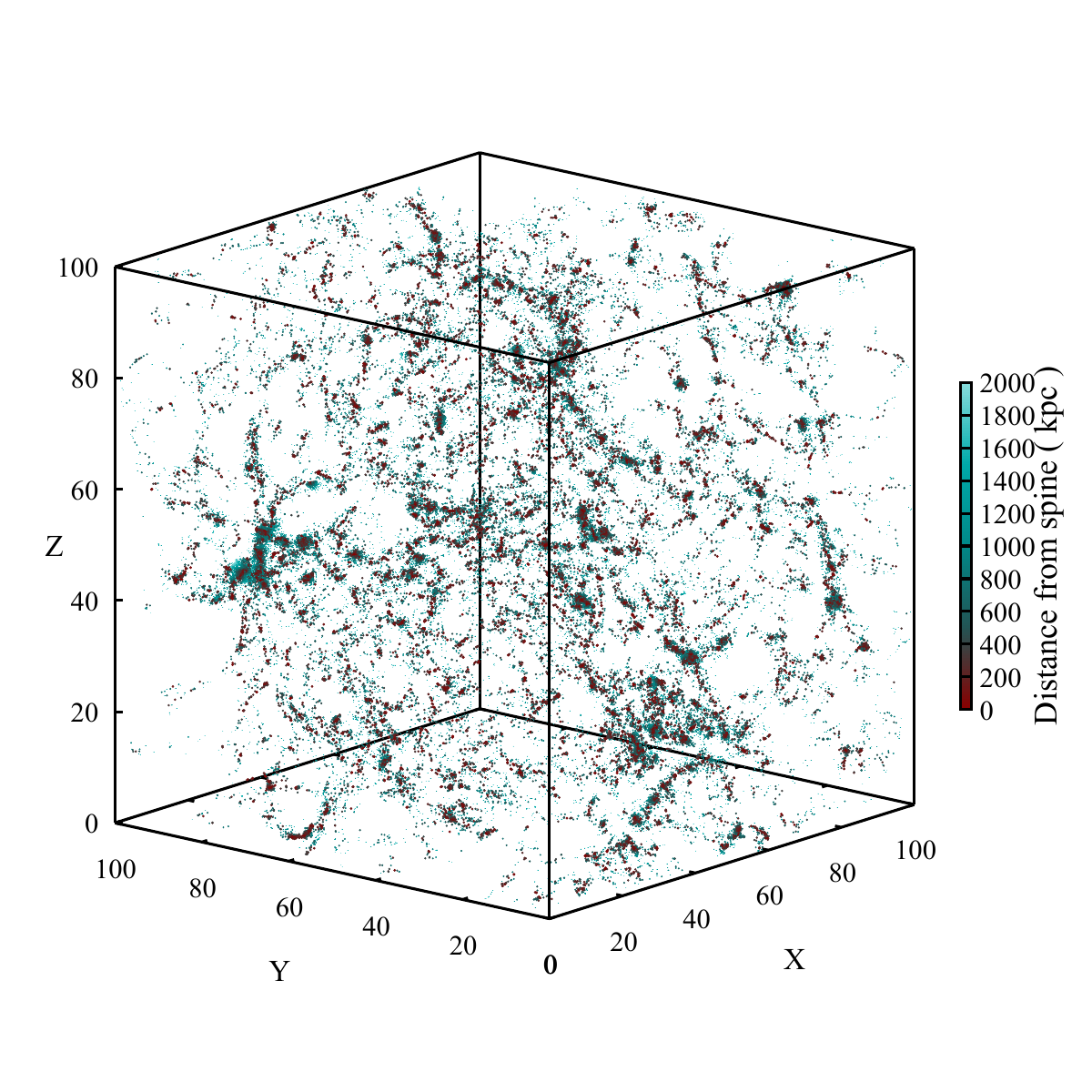}\\
\caption{In the left panel of this figure, we show the critical points
  identified by DisPerSE in one of the 100 realizations from EAGLE. In
  the right panel, we display the galaxies identified in the
  filamentary environment for the same realization.}
\label{fig:disp}
\end{figure*}

{}\begin{table*}{}
\centering
\begin{tabular}{|l|c|c|c|}
\hline
Linking length  & 0.5  Mpc & 1  Mpc  & 2  Mpc\\ 
\hline
Number of filaments including pairs &  $2217 \pm 19$ &  $1524 \pm 21$ & $170 \pm 8$ \\
\hline
\textcolor{black}{Number of filaments excluding pairs} &  \textcolor{black}{$883 \pm 12$} &  \textcolor{black}{$920 \pm 16$} & \textcolor{black}{$115 \pm 7$} \\
\hline
Number of pairs \textcolor{black}{in all filaments} &  $189648 \pm 620$ & $198141 \pm 630$ & $200726 \pm 610$\\ 
\hline
Number of isolated pairs \textcolor{black}{in all filaments} &  $997 \pm 21$ &  $2638 \pm 36$ &  $3348 \pm 40$\\ 
\hline
\end{tabular}
\caption{This table shows the number of filaments and number galaxy
  pairs hosted within the filaments for three different linking
  lengths. 1$\sigma$ deviations around the mean are obtained from 100
  realizations randomly drawn from the EAGLE data.}
\label{tab:counts}
\end{table*}

\subsection{Identification of filament groups} \label{subsec:filgroup}
\noindent First, we identify galaxies from EAGLE that reside within
filamentary environments using DisPerSE. Galaxies are classified as
being in a filamentary environment if they are located within $2 \mpc$
of the filament spine and are more than $3 \mpc$ away from the nearest
node. Once these galaxies are identified, we group them into filament
groups using the FoF algorithm \cite{huchra82, berlind06}. Each galaxy
maintains a list of neighboring galaxies known as its
\textit{friends}. Groups are formed by galaxies sharing the same
\textit{friend-list}. Selecting neighboring galaxies involves a
subjective decision guided by the \textit{linking length}
parameter. Typically, the linking length ($l_{L}$) is set to $0.2$
times the mean inter-particle separation ($\bar{r}_g$). The mean
inter-galactic separation of the filamentary galaxy distribution is
approximately $2.5 \mpc$. Initially, using $l_{L} = 0.2 \,\bar{r}_g$
resulted in filament groups significantly smaller than the actual span
between nodes. We carried out our analysis with different values of
$l_{L}$ including $0.5$, $1$ and $2 \mpc$ to ensure robustness of our
results. Subsequently, for the next part of analysis, we retained only
those groups containing at least 10 members. The objective here is to
determine the effective direction of the filament and analyze the
relative orientation of galaxy pairs within it.

\subsection{Determination of filament axis} \label{subsec:filax}
\noindent We calculate the moment of inertia tensor for a given group
of galaxies identified as a filamentary system using the formula
\begin{eqnarray}
    \mathbf{I} = \sum_{i}{ m_{i} \left[ \,||\,\mathbf{r}_{i}\,||^{2}\,\, \mathbf{E}_{3} \,-\, \mathbf{r}_{i} \otimes \mathbf{r}_{i} \,\right]}
\end{eqnarray}
where $m_{i}$ represents the stellar mass of the $i^{th}$ galaxy. The
position vector $\mathbf{r}_{i}$ for each galaxy is determined
relative to the center-of-mass (CoM) of the group, which serves as the
origin.  The term $\mathbf{E}_3$ denotes the $3 \times 3$ identity
matrix, and $\mathbf{r}_{i} \otimes \mathbf{r}_{i}$ signifies the
outer product of $\mathbf{r}_{i}$ with itself, resulting in a
matrix.

We diagonalize the moment of inertia tensor to determine its three
eigenvalues $\lambda_1$, $\lambda_2$, and $\lambda_3$ along with
corresponding eigenvectors $\hat{q}_1$, $\hat{q}_2$ and $\hat{q}_3$
that define the principal axes. The three eigenvalues are obtained as
roots of the characteristic equation,
\begin{eqnarray}
   \mathrm{det} \left[\,\, \mathbf{I} - \lambda \mathbf{E}_{3} \,\,\right] = 0 .
\end{eqnarray}
The eigenvector $\hat{q}_3$, corresponding to the smallest eigenvalue
$\lambda_3$, identifies the axis that defines direction of the longest
extent or the smallest moment of inertia. Thus, $\hat{q}_3$ represents
the principal direction of the filament.

\subsection{Identification of galaxy pairs within filaments and analyzing their orientations} \label{subsec:moi}
\noindent Let us begin by considering two galaxies within a
filamentary group, denoted by their stellar masses $m_{1}$ and
$m_{2}$, forming a pair. We consider all galaxy pairs for which the
pair separation $r_p<200$ kpc and the mass ratio
$\frac{m_{1}}{m_{2}} < 10$ where $m_{1}>m_{2}$.
%$\frac{{m}_{1}}{{m}_{2}}<10$​ where ${m}_{1}>{m}_{2}$.
It may be worthwhile to mention here that a galaxy can be part of
multiple pairs. The galaxy pairs in each filament may have any
possible orientation with respect to the filament axis.  For each
galaxy pair, we can calculate the angle between the unit vector
$\hat{r}_p$ associated with the pair separation and the unit vector
$\hat{q}_{3}$ defining the axis of the host filament. The angle will
be given by,
\begin{eqnarray}
    \theta = \cos^{-1} \left( \, \hat{r}_{p} \cdot \hat{q}_{3} \, \right)
\end{eqnarray}
We calculate the probability distribution \textcolor{black}{$p (|\cos
  \theta|)$} utilizing the \textcolor{black}{$|\cos \theta|$} values
obtained for all the pairs.

Some pairs within filaments can be part of galaxy groups, where the
orientation of the pairs is mainly influenced by the group environment. 
We are also interested in analyzing a subset of pairs
that are less likely to be affected by nearby galaxies. To achieve
this, we define ``isolated pairs'' as those where no other galaxies
exist within a distance of $300$ kpc from each member of the pair.

\begin{figure*}[ht!]%
\centering
\includegraphics[width=0.45\textwidth, trim=0.0cm 0.0cm 0.0cm 0.0cm, clip=true]{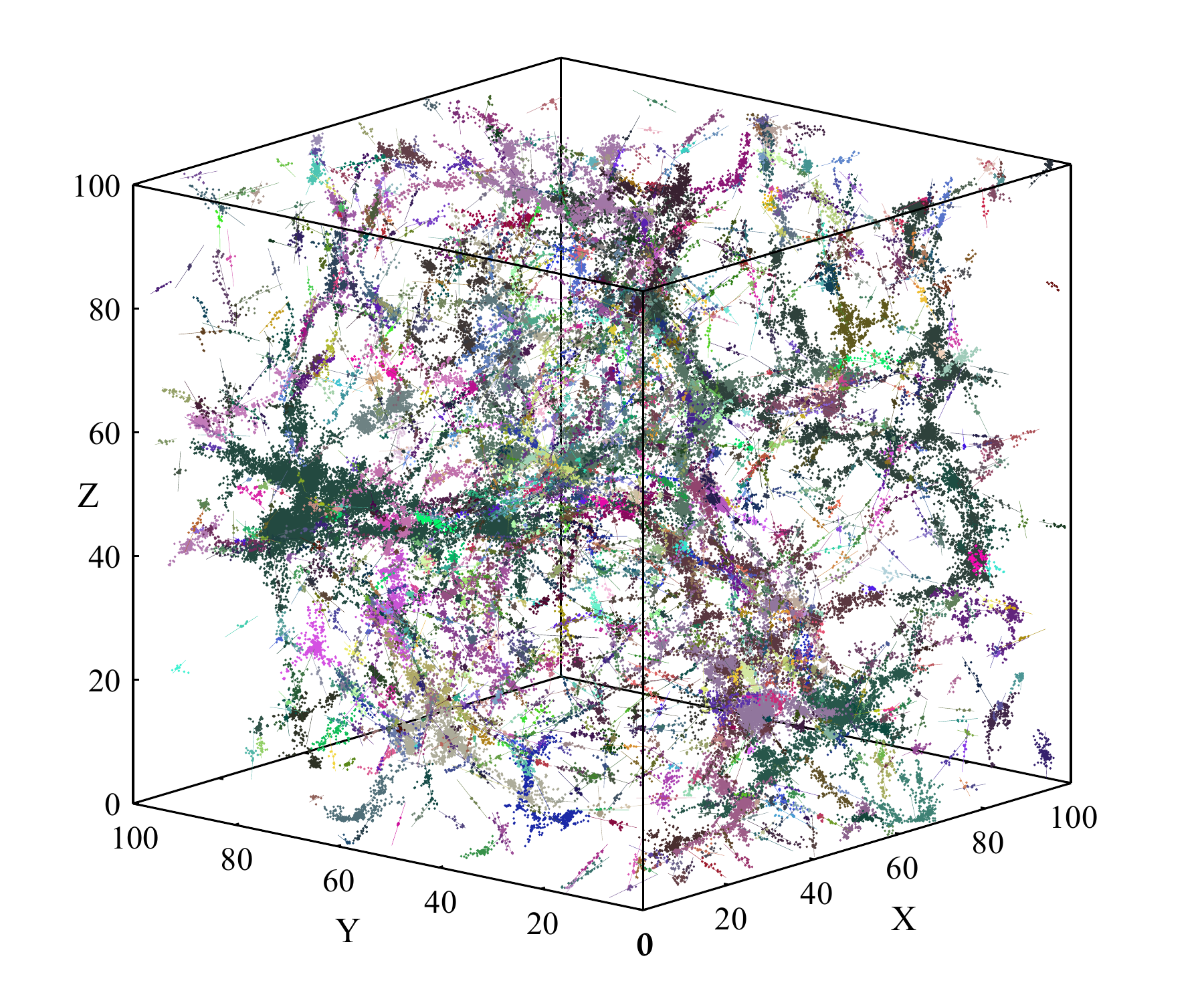}
\includegraphics[width=0.45\textwidth, trim=0.0cm 0.0cm 0.0cm 0.0cm, clip=true]{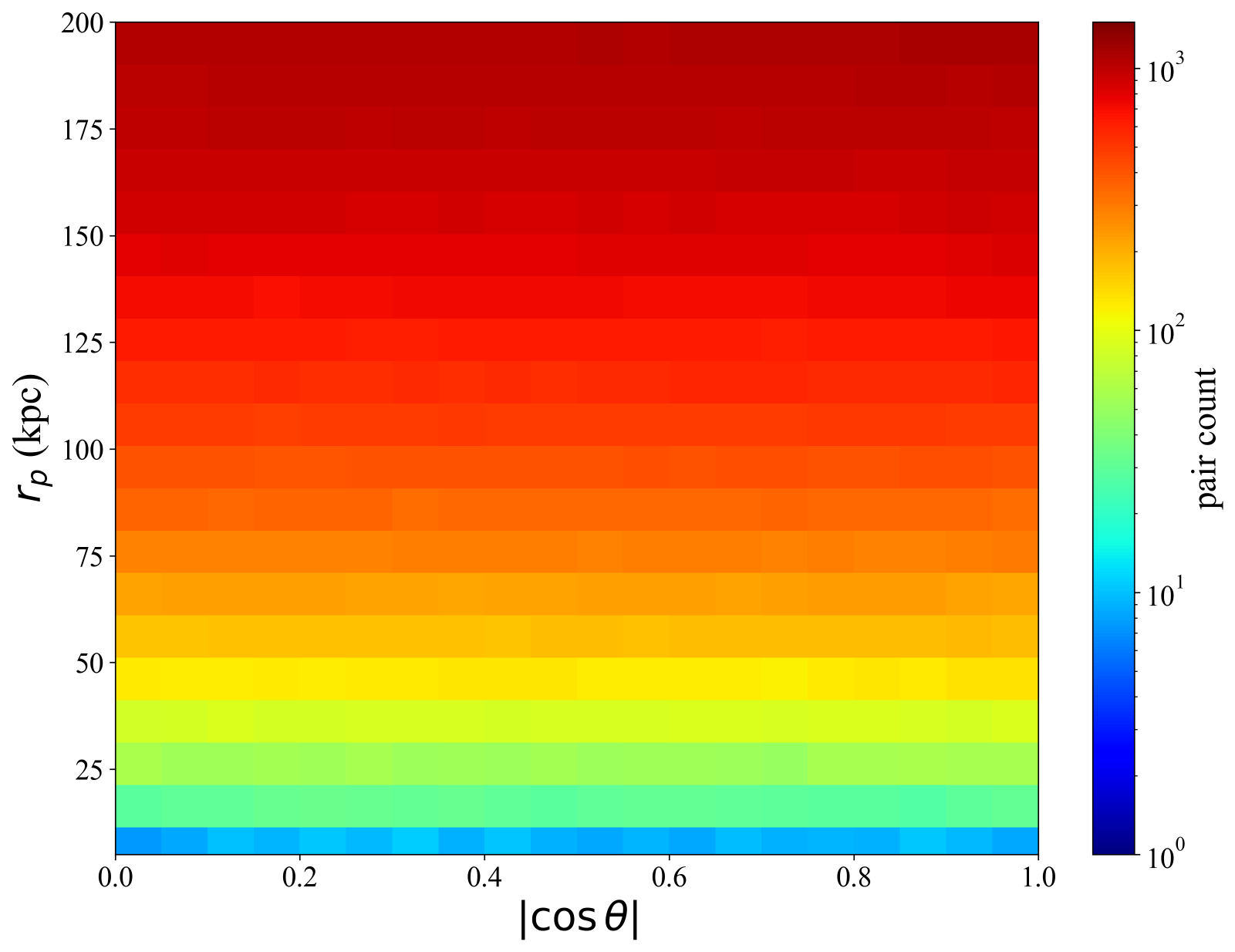}\\
\caption{In the left panel of this figure, filament groups identified
  by FoF using a linking length of $1 \mpc$ are displayed, with each
  filament axis represented by a straight line. The right panel shows
  the number of pairs available for various combinations of $r_p$ and
  \textcolor{black}{$|\cos \theta |$} for the same linking length. Here
  $\theta$ is the angle between any galaxy pair and its host
  filament.}
\label{fig:pair_orient}
\end{figure*}

\begin{figure*}[ht!]%
\centering
\includegraphics[width=\textwidth, trim=0.0cm 0.0cm 0.0cm 0.0cm, clip=true]{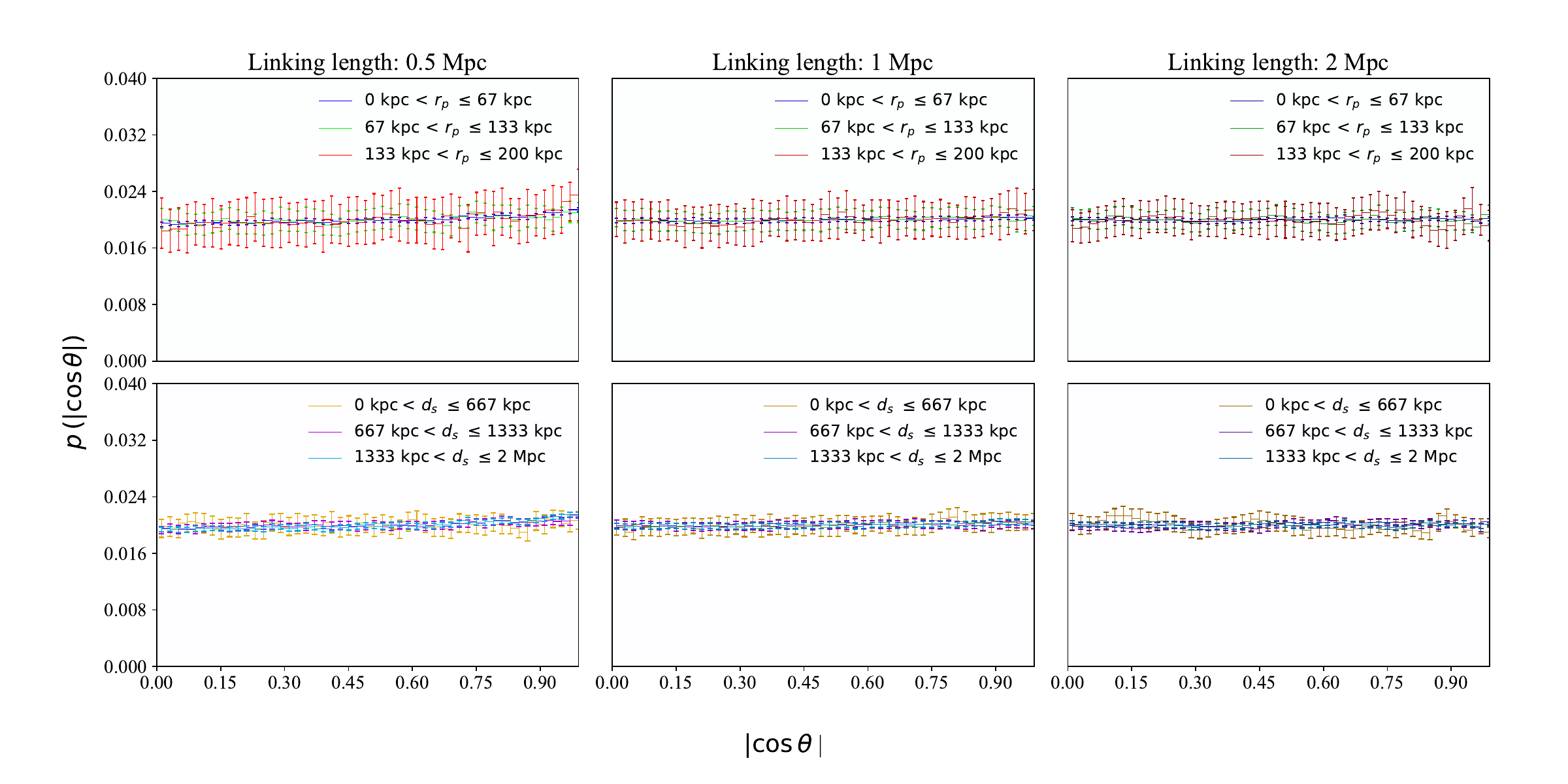}
\caption{The different panels of this figure show the probability
  distribution function (PDF) of \textcolor{black}{$|\cos \theta |$} for
  different $r_p$ (top) and $d_s$ (bottom) bins. The 1$\sigma$
  errorbars shown here are obtained from \textcolor{black}{100 random
    realizations drawn from the EAGLE simulation.}  The PDFs are shown
  for three different linking lengths $0.5 \mpc$ (left), $1 \mpc$
  (middle) and $2 \mpc$ (right). }
\label{fig:ll_dep}
\end{figure*}

\begin{figure*}[ht!]%
\centering
\includegraphics[width=\textwidth, trim=0.0cm 0.0cm 0.0cm 0.0cm, clip=true]{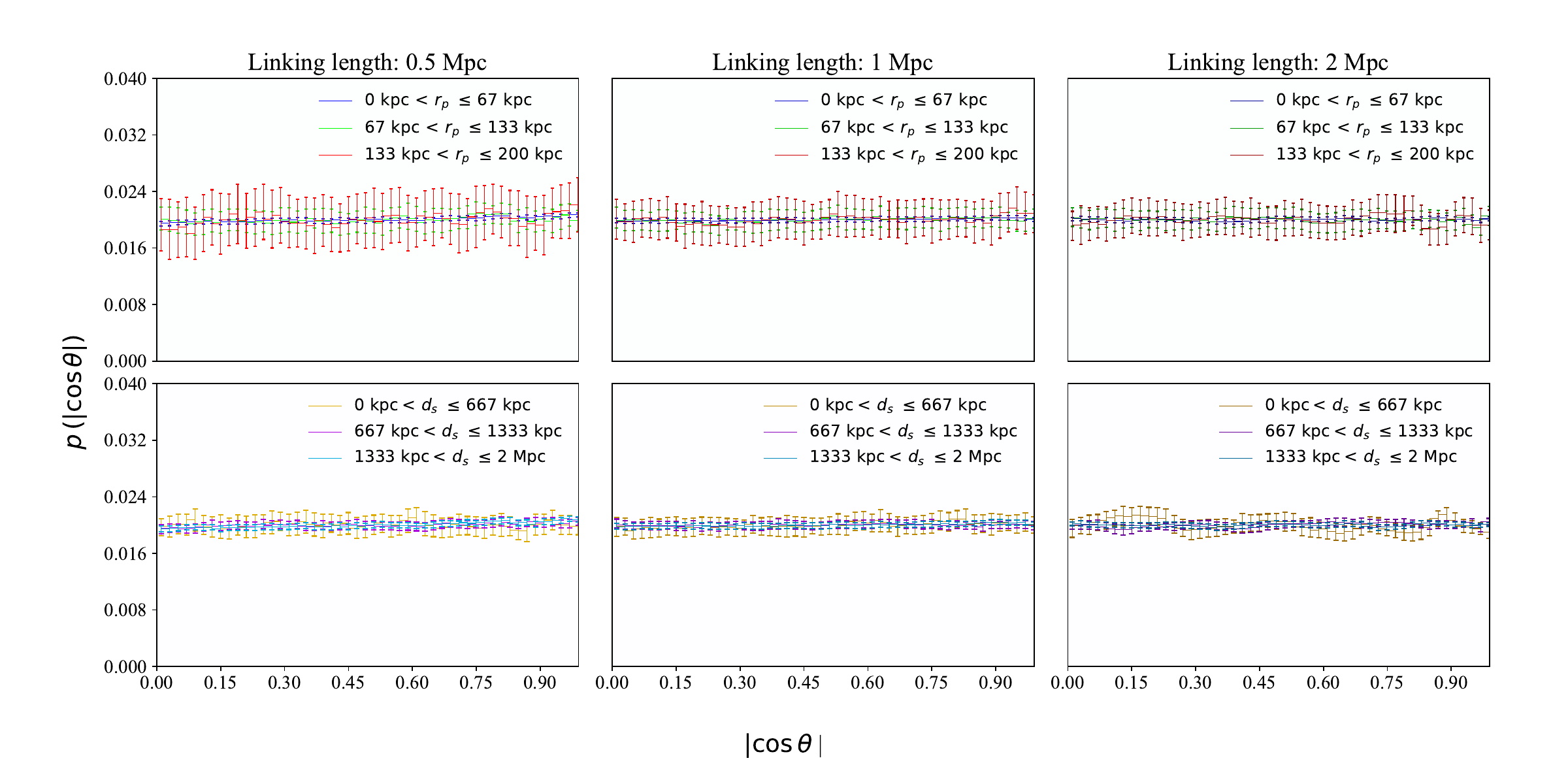}
\caption{Same as \autoref{fig:ll_dep2} but for the case where the
  galaxy pairs are excluded before finding the orientation of the
  filaments.}
\label{fig:ll_dep2}
\end{figure*}

\begin{figure*}[ht!]%
\centering
\includegraphics[width=\textwidth, trim=0.0cm 0.0cm 0.0cm 0.0cm, clip=true]{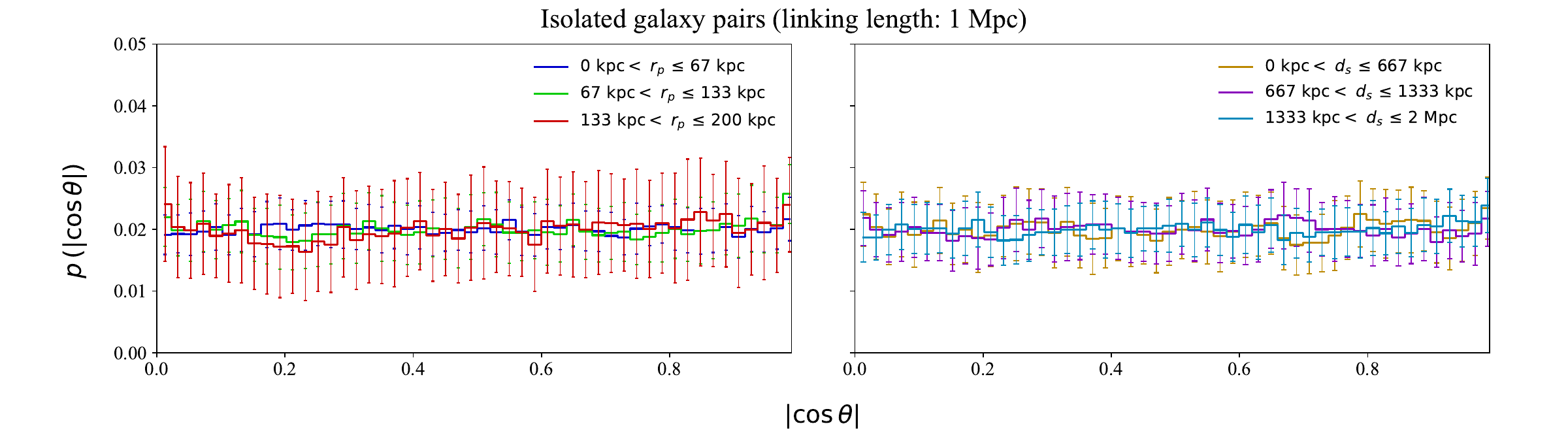}
\caption{Different panels of this figures show the PDF of
  \textcolor{black}{$|\cos \theta |$} in different $r_p$ and $d_s$ bins
  for the isolated pairs. The 1$\sigma$ error bars shown in each case
  are derived from \textcolor{black}{100 random realizations drawn from
    the EAGLE simulation.}}
\label{fig:isol_orient}
\end{figure*}

\begin{figure*}[ht!]%
\centering
\includegraphics[width=\textwidth, trim=0.0cm 0.0cm 0.0cm 0.0cm, clip=true]{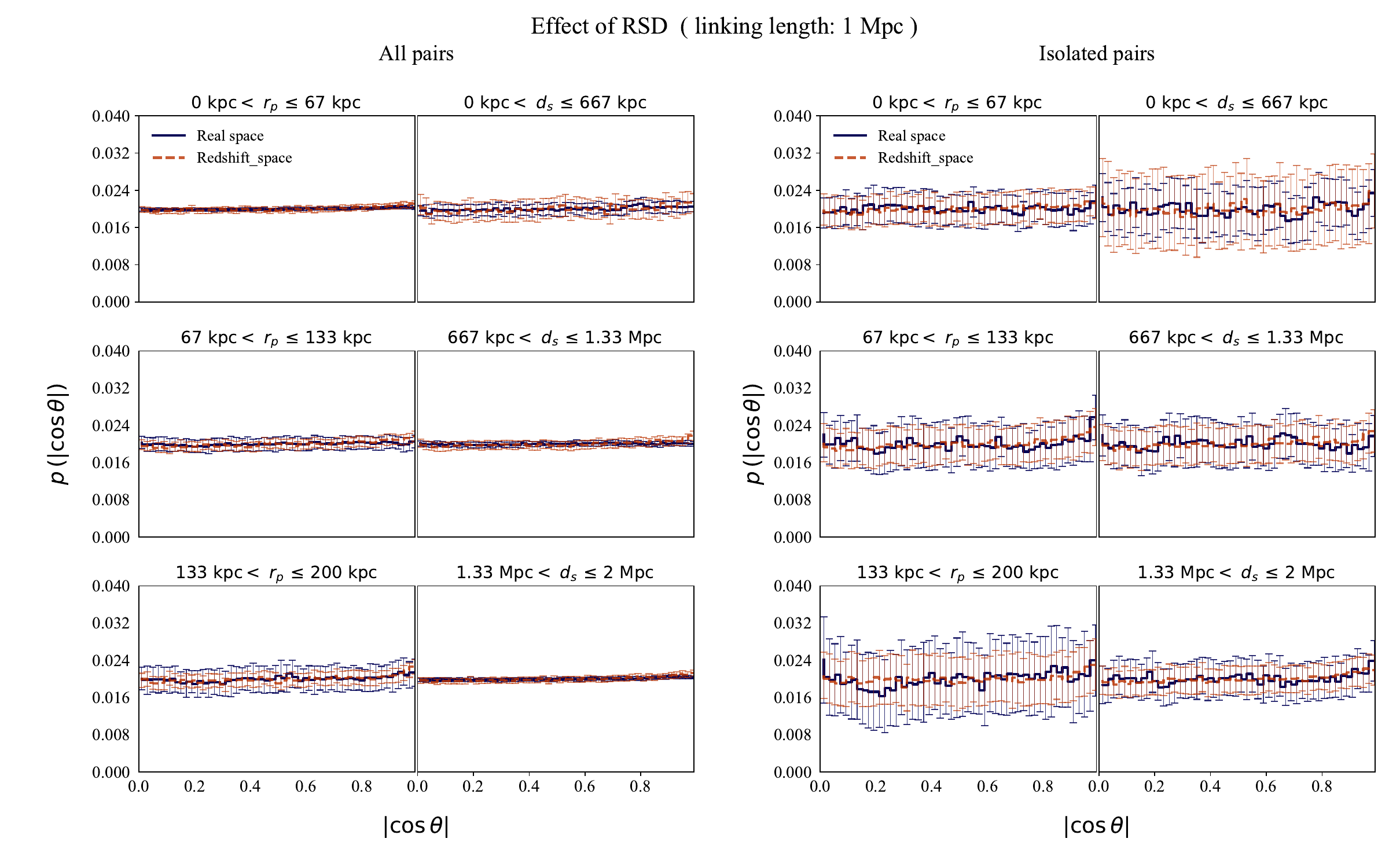}
\caption{ This figure shows the PDF of \textcolor{black}{$| \cos \theta
    |$} estimated in real and redshift space for linking length $1
  \mpc$ are compared for different $r_p$ and $d_s$ bins. 1$\sigma$
  error bars are drawn from \textcolor{black}{100 random realizations
    drawn from the EAGLE simulation.} in each case. The
  \textcolor{black}{$p(|\cos \theta|)$} for all the available pairs are
  shown on the left. The same is shown for the isolated pairs on
  right.  }
\label{fig:rsd_effect}
\end{figure*}

\section{Results} \label{sec:results}
In the left panel of \autoref{fig:disp}, we display critical points
identified by DisPerSE from one randomly sampled realization of the
EAGLE simulation. Galaxies located within a distance $d_s \leq 1$ Mpc
from the filament spine and farther than $d_n > 2$ Mpc from the nodes
are identified as part of filamentary environments. These galaxies are
collectively shown in the right panel of \autoref{fig:disp} for the
same realization. To analyze individual filaments, we apply the FoF
algorithm to datasets comprising galaxies within filamentary
environments. The filamentary groups, identified using a linking
length of $1 \mpc$ from the same realization of EAGLE data are shown
in the left panel of \autoref{fig:pair_orient}. The axis of each
filamentary group is represented by a straight line in this figure. We
isolate galaxy pairs within each filamentary group and measure the
angle between the pair separation vector and the axis of the host
filament for all such pairs.

The 2D histogram in the right panel of \autoref{fig:pair_orient} shows
the number of available galaxy pairs in filamentary environments for
different combinations of pair separation $r_p$ and
\textcolor{black}{$|\cos \theta|$}.  Here $\theta$ denotes the angle
between the pair separation vector and the axis of the filament
hosting the pair. \textcolor{black}{The plot shows that the number of
  galaxy pairs at any given separation ($r_p$) is nearly independent
  of the orientation of the pairs relative to the axis of their host
  filaments. A gradient along the $r_p$ axis indicates that the number
  of available pairs gradually increases with separation, for $r_p
  \leq 200$ kpc. The results in this figure do not suggest any
  preferential alignment of galaxy pairs within cosmic filaments.}

The FoF algorithm employed to identify distinct filaments relies on a
user-defined parameter known as the linking length. Varying this
parameter yields different sets of identified
filaments. \textcolor{black}{Therefore, it is crucial to repeat the
  analysis using various linking lengths.} To address this, we conduct
our analysis using three different linking lengths: $0.5$ Mpc, $1$
Mpc, and $2$ Mpc. The probability distributions of
\textcolor{black}{$|\cos \theta|$} for these linking lengths are
presented in the left, middle, and right panels of
\autoref{fig:ll_dep}.

\textcolor{black}{Any potential alignment signal for galaxy pairs is
  expected to depend on both the pair separation ($r_p$) and the
  distance of the pairs from the filament spine ($d_s$). With this in
  mind, we examine the orientation of galaxy pairs within filaments
  across different ranges of $r_p$ and $d_s$.}

For a linking length of $0.5$ Mpc, the probability distributions of
\textcolor{black}{$|\cos \theta|$} for distinct intervals of $r_p$ and
$d_s$ are respectively shown in the top and bottom left panels of
\autoref{fig:ll_dep}. \textcolor{black}{$100$ bins are used to calculate
  the probability distribution of $\cos \theta$ in each case and
  throughout the rest of our analysis}. \textcolor{black}{Results from
  the top and bottom left panels indicate a nearly uniform probability
  distribution for each range of $r_p$ and $d_s$. This further
  suggests that galaxy pairs do not exhibit any preferential alignment
  relative to filament axes, regardless of their distance from the
  filament spine and for any pair separation.}

Results for the other two linking lengths, $1$ Mpc and $2$ Mpc, are
displayed in the middle and right panels of
\autoref{fig:ll_dep}. Interestingly, we observe nearly identical
probability distributions of \textcolor{black}{$|\cos \theta|$} across
all three linking lengths considered in our
analysis. 

\textcolor{black}{To measure the orientation of the filaments
  independently, we exclude the galaxy pairs when assessing filament
  orientation. Only the filaments containing at least 10 galaxies,
  after excluding paired galaxies, are considered. We then repeat our
  measurements of the galaxy pair orientations using these filament
  axes, and present the results in \autoref{fig:ll_dep2}. The observed
  trend is consistent with that in \autoref{fig:ll_dep}. We show the
  number of filaments and galaxy pairs for each linking length in
  \autoref{tab:counts}. The table also lists the number of available
  filaments with at least $10$ galaxies after excluding the galaxy
  pairs.}

\textcolor{black}{The galaxy pairs may belong to groups residing within
  a filament, where local conditions could influence the orientation
  of galaxy pairs relative to the filament spine. To investigate this,
  we repeat our analysis focusing on isolated galaxy pairs and present
  the results for a specific linking length ($l_{L}=1$ Mpc) in
  \autoref{fig:isol_orient}. The same number of bins was used to
  compute the probability distributions of \textcolor{black}{$|\cos
    \theta|$}. Our findings again show a nearly uniform probability
  distribution across different ranges of $r_p$ and $d_s$, as shown in
  the two panels of \autoref{fig:isol_orient}.}

We conduct our analysis using EAGLE data in real space. However,
observations from various redshift surveys map galaxies in redshift
space, where peculiar velocities introduce distortions in the
clustering pattern. It is crucial to examine the impact of these
redshift space distortions (RSD) on the anticipated alignment
signal. We map the positions of EAGLE galaxies in redshift space
considering their peculiar velocities, and then repeat our analysis to
assess the influence of RSD on our findings.

If $\mathbf{v}_{i}$ and $\mathbf{r}_{i}$ represent the velocity
and real-space position of the $i^{th}$ galaxy relative to an
observer, then the redshifted position $\mathbf{s}_{i}$ as measured
by the observer can be estimated as,
\begin{eqnarray}
    \mathbf{s}_{i} = \mathbf{r}_{i} \, \left[\, 1 +
      \frac{\mathbf{v}_{i} \cdot \,\mathbf{r}_{i}}{H_0\,\,
        ||\,\mathbf{r}_{i}\,||^{2} }\, \right].
\end{eqnarray}
where $H_0$ is the present value of the Hubble parameter. The 100 Mpc
box from EAGLE simulation represents the galaxy distribution at
$z=0$. From the EAGLE data, we generate 100 mock redshift space
distributions by randomly placing the observer at one of the 8 corners
each time. From these redshifted distributions, we extract the same
number of galaxies as used in the real space analysis. We then follow
the exact steps outlined in Section \ref{sec:methods} to analyze the
orientation of galaxy pairs within filaments extracted from these
redshift space distributions.

\textcolor{black}{Some of the filaments detected in redshift space may
  result from the ``Fingers of God'' (FOG) effect. The orientation of
  galaxy pairs within these spurious filaments could influence the
  probability distributions of $|\cos \theta|$. We show the
  probability distributions of $|\cos \theta|$ for all pairs and
  isolated pairs in different panels of \autoref{fig:rsd_effect}. For
  each type of pair, we compare the results in real space and redshift
  space across various ranges of $r_p$ and $d_s$ in separate panels of
  \autoref{fig:rsd_effect}. Our findings clearly demonstrate that,
  despite the presence of FOGs, the probability distributions of the
  cosine of the orientation angles remain uniform in redshift space
  for each range of $r_p$ and $d_s$. In each panel, the 1$\sigma$
  error bars for all pairs are comparable in real and redshift
  space. However, the error bars for the isolated pairs are relatively
  larger due to the smaller number of pairs in these samples.}

\section{Conclusions}
We study 3D orientation of galaxy pairs within filaments using data
from EAGLE simulation. We carried out our analysis for different
subsets of filaments identified using 3 different linking lengths. We
also consider the isolated pairs that are less impacted by their
immediate neighbourhood. In each case, we study the orientations for
galaxy pairs with different pair separation ($r_p$) and distance from
filament spine ($d_s$). For each linking length, $r_p$ and $d_s$, we
recover a nearly identical probability distribution of
\textcolor{black}{$|\cos \theta|$} where $\theta$ is the angle between
galaxy pair and axis of filament.  \textcolor{black}{We validate our
  method using controlled simulations as described in the Appendix
  (\autoref{sec:validate}). Our results are consistent with a random
  orientation of galaxy pairs within filaments.}

\textcolor{black}{The role of the large-scale environment in galaxy
  evolution remains a debated topic in the current literature. Our
  analysis shows that galaxy pairs do not exhibit any preferential
  alignment within filaments, both in real and redshift space. We
  observe a nearly uniform probability distribution of the cosine of
  the orientation angle for each range of $r_p$ and $d_s$ considered
  in our study. The size of the error bars differs between real and
  redshift space, primarily due to the presence of FOGs and the
  apparent elongation and compression of structures along the line of
  sight.}

\textcolor{black}{Studying the three-dimensional orientations of galaxy
  pairs relative to filament axes using observational datasets
  presents several challenges. Some earlier studies \citep{tempel15a,
    mesa18} with SDSS data projected both filaments and galaxy pairs
  onto the plane of the sky, revealing a $15\%-25\%$ excess of aligned
  pairs compared to a random distribution. In \citep{tempel15a},
  filaments are identified using the Bisous model, which is based on a
  marked point process \citep{tempel14}. To avoid correlations between
  filament and galaxy pair orientations, they replace each galaxy pair
  with a single point during filament extraction. We carry out our
  analysis by both including and excluding galaxy pairs when
  evaluating the filament orientation.}

\textcolor{black}{ \cite{tempel15a} measure the alignment signal by
  projecting galaxy pairs and filaments onto the plane of the sky,
  where the orientation angle is constrained to the range of $0$ to
  $90$ degrees. In contrast, our analysis is performed in three
  dimensions (3D), where the orientation angle can vary between $0$
  and $180$ degrees. The earlier analysis primarily focus on measuring
  the alignment signal using observational datasets whereas our
  analysis only deals with the data from hydrodynamical
  simulation. Extending the interpretation of the results in previous
  analyses to 3D is not straightforward due to the projection effects
  and redshift space distortions. We cannot directly compare our
  results with those from earlier studies due to the differences in
  the applied methodologies. However, it would be interesting to
  explore the impact of different filament identification strategies
  on potential alignment signals. Some caveats in our analysis are
  related to the identification of the filaments using
  DisPerSE. DisPerSE identifies structures based on their
  ``persistence'', which reflects how long a structure persists across
  different smoothing scales. The choice of persistence threshold can
  significantly affect the resulting filament spines. Higher
  thresholds may lead to missing smaller or weaker filaments, while a
  very low threshold can result in noisy structures or false
  positives. The choice of smoothing scale and number of smoothing can
  also influence how the filaments are traced. Different choices of
  these parameters may lead to different sets of filaments. Further
  studies with different filament identifications techniques are
  necessary before arriving at a definite conclusion on this issue.}

\textcolor{black}{The primary objective of our study is to perform a 3D
  analysis of galaxy pair alignment within cosmic filaments using data
  from a hydrodynamical simulation. This represents the first 3D
  analysis of galaxy pair alignment in cosmic filaments, and our
  findings show no evidence of alignment signals in the
  simulations. It would be interesting to carry out a similar 3D
  analysis using observational data to investigate whether galaxy pair
  alignment signals exist in the real universe. We plan to pursue this
  investigation in future work.}

\section*{Data availability}
Data used for this study is publicly available in the database of the
EAGLE project. Data generated through this project can be shared upon
reasonable request to the authors.

\section*{Acknowledgement}
The authors thank the anonymous reviewer for his insightful suggestions and comments. SS thanks DST-SERB for support through the NPDF project
PDF/2022/000149.  SS also thanks Prof. Somnath Bharadwaj for useful discussions. BP would like to acknowledge financial support from the SERB, DST, Government of India through the project CRG/2019/001110.
BP would also like to acknowledge IUCAA, Pune for providing support through associateship programme.  The authors acknowledge the Virgo
Consortium for making their simulation data publicly available.  The
EAGLE simulations were performed using the DiRAC-2 facility at Durham,
 managed by the ICC, and the PRACE facility Curie based in France at
TGCC, CEA, Bruy\`{e}res-le-Ch\^{a}tel.

\bibliographystyle{JHEP}
\bibliography{orient.bib}
\appendix 

\section{Validating the method using controlled simulations}\label{sec:A1}
\label{sec:validate}
\begin{figure*}[ht!]%
\centering \includegraphics[width=0.9\textwidth, trim=0.0cm 0.0cm
  0.0cm 0.0cm, clip=true]{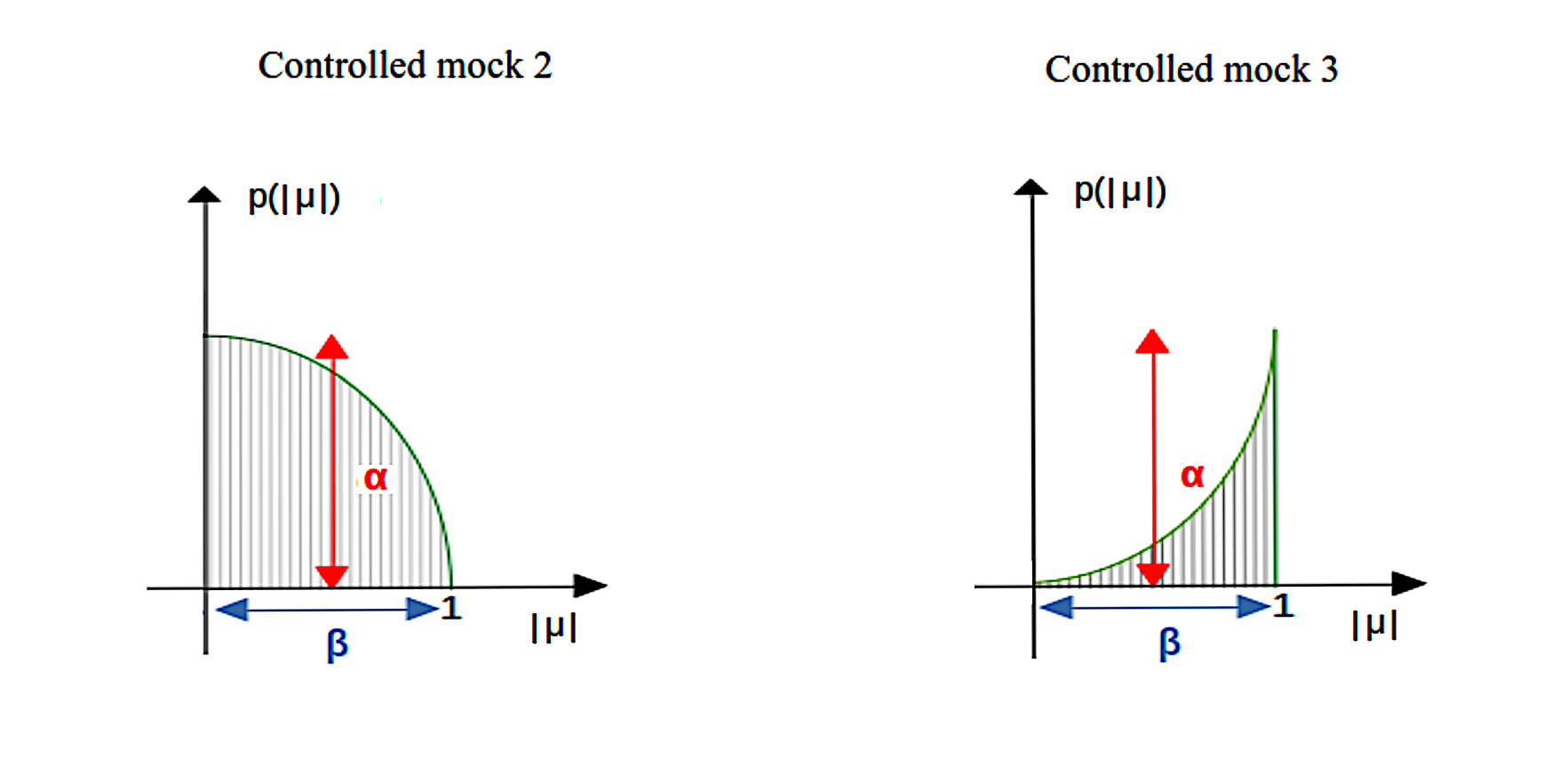}
\caption{Schematic diagram showing the probability distributions of
  Controlled mock 2 and 3}
\label{fig:CM}
\end{figure*}

\noindent In this section, we assess the reliability of our method by
conducting tests with controlled simulations. We create multiple
controlled simulations of galaxy pairs within mock filaments using
Monte Carlo simulations. \textcolor{black}{ We randomly generate the
  axis of each filament, and then position the galaxy pairs around the
  filament axis. The galaxy paires are assignined orientation angles
  according to a specific probability distribution of $\cos \theta$,
  where $\theta$ represents the orientation angle.} We examine three
distinct types of probability distributions which are discussed
below.\\
\begin{itemize}
\item {Controlled mock 1: We simulate pairs with random orientation
  relative to the filament axis using a discrete probability
  distribution $p(\cos\theta) = \frac{1}{N_b}$, where $N_b$ represents
  the number of bins used for binning $\cos \theta$. When orientations
  are random, the probability of any specific orientation of galaxy
  pairs relative to the filament axis follows a uniform distribution
  given by $p(\mu) = \frac{1}{(\mu_{max}-\mu_{min})}$ where $\mu=\cos
  \theta$ and $\mu_{min}=-1$ and $\mu_{max}=1$ are the minimum and
  maximum values of $\mu$. The discrete version of this distribution
  is derived by multiplying the continuous probability distribution by
  the bin width $\Delta \mu=\frac{\mu_{max}-\mu_{min}}{N_b}$. Hence,
  the discrete probability distribution for Controlled mock 1 is
  expressed as,
\begin{eqnarray}
    p(\cos \theta) = \frac{1}{N_b} \nonumber
    \end{eqnarray}}

  \item {Controlled mock 2: The probability distribution for pairs
    with preferential orientation perpendicular to the filament axis
    can be generated using the distribution $p(\cos \theta) =
    \frac{4}{\pi N_b} \sqrt{1- \cos^2 \theta} $. The distribution can
    be derived as follows. In the left panel of Figure \ref{fig:CM},
    we illustrate the total probability for Controlled mock 2,
    represented by the shaded area which, by definition, must be equal
    unity. Here, we denote $\cos \theta$ as $\mu$. If $p(\mu)$ has its
    maximum value $\alpha$ and $\mu$ spans from $\mu_{min}=-\beta$ to
    $\mu_{max}=\beta$, the area of the shaded region would be
    $\frac{1}{2}\pi \alpha \beta$. So we have $\alpha = \frac{2}{\pi}$
    as $\beta=1$. The equation of this ellipse becomes
 \begin{eqnarray}
    \frac{\mu^2}{\beta^2} + \frac{p(\mu)^2}{\alpha^2} = 1
    \end{eqnarray}
The probability distribution of $\mu$ i.e. $\cos \theta$ can be written as
    \begin{eqnarray}
    p(\cos \theta) = \frac{2}{ \pi} \sqrt{1-\cos^2 \theta} \nonumber
    \end{eqnarray}

 Thus, the discrete version of the probability distribution for
Controlled mock 2 is given by
 \begin{eqnarray}
    p(\cos \theta) = \frac{4}{ \pi N_b} \sqrt{1-\cos^2 \theta} \nonumber
    \end{eqnarray}
}

  \item {Controlled mock 3: The pairs exhibiting preferential
    alignment along the filament axis are governed by the distribution
    $p(\cos \theta) = \frac{4}{(4-\pi)N_b} \left[ 1 - \sqrt{1 - \cos^2
        \theta} \right]$. In the right panel of Figure \ref{fig:CM},
    we present the probability distribution for Controlled mock 3. By
    definition, the shaded area in this figure must equate to unity,
    leading to the relationship $2\alpha \beta - \frac{1}{2}\pi \alpha
    \beta =1$, where $\alpha = \frac{2}{4 - \pi}$ denotes the maximum
    probability for parallel orientation. The equation of the ellipse
    describing this scenario is given by
\begin{eqnarray}
    \frac{\mu^2}{\beta^2} + \frac{\left[ \alpha - p(\mu) \right]^2}{\alpha^2} = 1
\end{eqnarray}
Therefore, considering $N_b$ discrete bins, the discrete version of
the probability distribution is expressed as
\begin{eqnarray}
    p(\cos \theta) = \frac{4}{ (4 - \pi )N_b} \left[ 1- \sqrt{1-\cos^2
        \theta} \right] \nonumber
\end{eqnarray}
}
\end{itemize}

\begin{figure*}[ht!]%
\centering \includegraphics[width=\textwidth, trim=0.0cm 0.0cm 0.0cm
  0.0cm, clip=true]{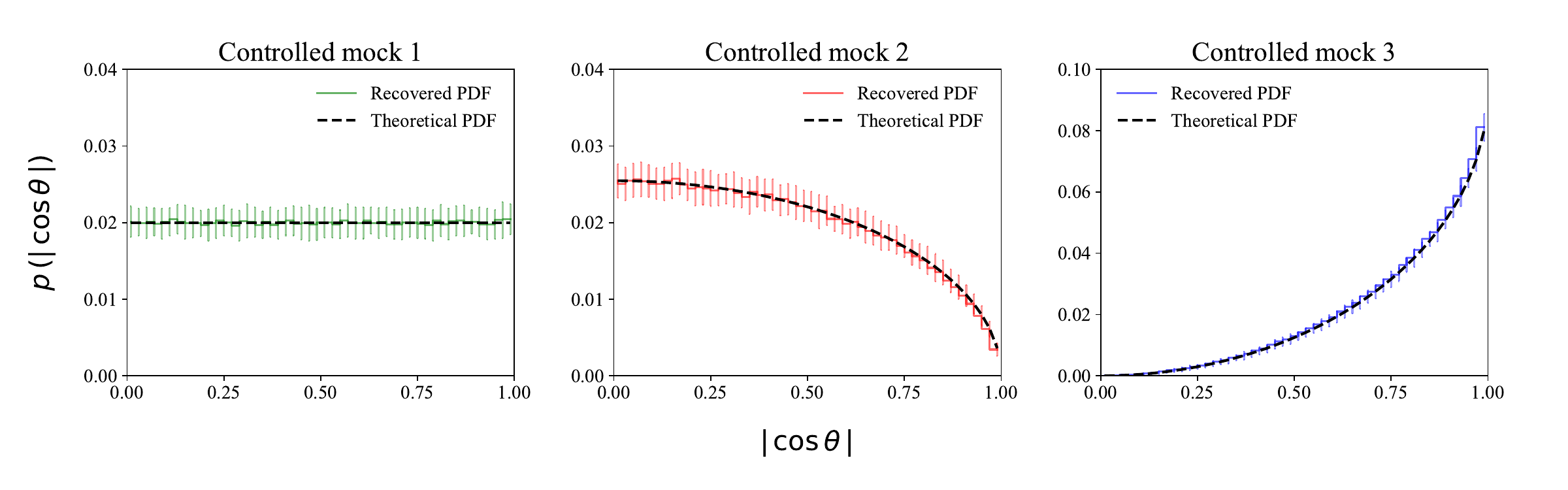}
\caption{This figure shows the recovered and the theoretical
  probability distributions for the three different types of
  controlled mock samples. The 1$\sigma$ errorbars in each case are
  estimated using 100 Monte-Carlo realizations. \textcolor{black}{The
    theoretical probability distribution of $p(\cos \theta)$ in each
    case is multiplied by 2 since we present the probability
    distributions of $p(|\cos \theta)|$ in these plots.}}
\label{fig:A1}
\end{figure*}

\begin{table*}{}
\centering
\label{tab:mock_gen}
\begin{tabular}{ll}
\hline
Box size & $100 \mpc$    \\   
Filament length& $30 \mpc$      \\
Number of pairs & $5000$  \\   
Maximum pair separation & $200 \,\,\mathrm{kpc}$ \\
Max distance from filament spine & $2 \mpc$\\
\hline
\end{tabular}
\caption{This table shows the parameters used in the controlled
  simulations of the mock filaments and galaxy pairs in these tests.}
\label{tab:sim}
\end{table*}

\noindent We use ${N_b} = 100$ throughout this analysis. We ensure
$\displaystyle \sum_{i=1}^{N_b} p(|\cos \theta_i|) = 1$, for each of
the three cases, where $\theta_i$ represents the orientation angle at
the center of the $i^{th}$ bin. We generate 100 mock samples in each
case to estimate the 1$\sigma$ errorbars.

We attempt to reconstruct the probability distribution of galaxy pairs
in these three types of mock samples using the methodologies described
in Section \ref{sec:methods}. For each mock sample, we treat the
points associated with the entire set of pairs from a single
filament. Using the technique outlined in Section \ref{subsec:moi},
we determine the Center of Mass (CoM) and the axis of the
filament. Subsequently, we calculate the angle of each pair relative
to the identified filament axis.

The reconstructed probability distributions of the orientation of
galaxy pairs, along with their corresponding theoretical
distributions, are displayed in separate panels of Figure
\ref{fig:A1}. We also conducted repeated tests where we varied the
parameters of the simulated filaments listed in \autoref{tab:sim}. In
every instance, we consistently recovered the theoretical probability
distributions.

We observe that our method reliably recovers the theoretical probability 
distribution in each scenario. \textcolor{black}{We also repeat our analysis 
by varying the length, diameter, and orientation of the simulated filaments, 
and find that our results are  independent of these factors. However, 
it is worth mentioning that the simulated filaments are straight with a 
cylindrical geometry. They have finite length and diameter as listed in 
\autoref{tab:sim}. These filaments do not have the same complex or curved 
geometry as those identified in the EAGLE simulation, nor are they identified 
using DisPerSE. As a result, filament identification mismatches are not 
included in this analysis.}
\label{lastpage}
\end{document}